\documentclass[]{interact}

\usepackage{epstopdf}
\usepackage{subfigure}

\usepackage{natbib}
\bibpunct[, ]{(}{)}{,}{a}{}{,}
\renewcommand\bibfont{\fontsize{10}{12}\selectfont}

\theoremstyle{plain}

\theoremstyle{definition}

\theoremstyle{remark}

\usepackage{algorithm}
\usepackage{xcolor}
\usepackage{algorithmic}

\begin{document}

\articletype{RESEARCH ARTICLE}

\title{A data relocation approach for terrain surface analysis on multi-GPU systems: a case study on the total viewshed problem}

\author{
\name{A.~J. Sanchez-Fernandez\textsuperscript{a}\thanks{CONTACT A.~J. Sanchez-Fernandez. Email: sfandres@uma.es}, L.~F. Romero\textsuperscript{a}, G. Bandera\textsuperscript{a} and S. Tabik\textsuperscript{b}}
\affil{\textsuperscript{a}Department of Computer Architecture, University of Malaga, 29071 Malaga, Spain; \textsuperscript{b}Andalusian Research Institute in Data Science and Computational Intelligence, University of Granada, 18071 Granada, Spain}
}

\maketitle

\begin{abstract}
Digital Elevation Models (DEMs) are important datasets for modelling the line of sight, such as radio signals, sound waves and human vision. These are commonly analyzed using rotational sweep algorithms. However, such algorithms require large numbers of memory accesses to 2D arrays which, despite being regular, result in poor data locality in memory. Here, we propose a new methodology called skewed Digital Elevation Model (sDEM), which substantially improves the locality of memory accesses and increases the inherent parallelism involved in the computation of rotational sweep-based algorithms. In particular, sDEM applies a data restructuring technique before accessing the memory and performing the computation. To demonstrate the high efficiency of sDEM, we use the problem of total viewshed computation as a case study considering different implementations for single-core, multi-core, single-GPU and multi-GPU platforms. We conducted two experiments to compare sDEM with (i) the most commonly used geographic information systems (GIS) software and (ii) the state-of-the-art algorithm. In the first experiment, sDEM is on average 8.8x faster than current GIS software despite being able to consider only few points because of their limitations. In the second experiment, sDEM is 827.3x faster than the state-of-the-art algorithm in the best case.
\end{abstract}


\section{Introduction}
There are many problems of terrain surface analysis which require the evaluation of the data around a reference point. This is the case in viewshed computation where the reference point on the Digital Elevation Model (DEM) is usually called point of view (POV). This topic has been thoroughly studied in the recent literature~\citep{wang2019fast,dou2019equal}, being usually addressed using rotational plane sweep-based algorithms, or only rotational sweep~\citep{choset2005principles}. In particular, a line is traced from the POV, which works as a vertex in the plane. This line is rotated by $2\pi$ radians and all the points that cross that line are analyzed with respect to the vertex. Another related approach involves the discretization of the plane in azimuthal sectors radiating from the reference point~\citep{tabik2014efficient}. Every azimuthal sector is represented by an axis placed in its centre, and points crossed over by the axis are compared to the reference location. In this approach, the statistical representativeness of every axis progressively decreases as we move away from the vertex since the width of the sector increases linearly with the radius. However, in most cases, the required accuracy will also be reduced to the same extent. This issue is exploited in some situations where the reference location works as the transmitter or receiver of certain signals whose strength decreases with the square of the distance, such as radio signals, sound waves, and line of sight. 

\begin{figure}[b!]
\centering
\includegraphics[width=\columnwidth]{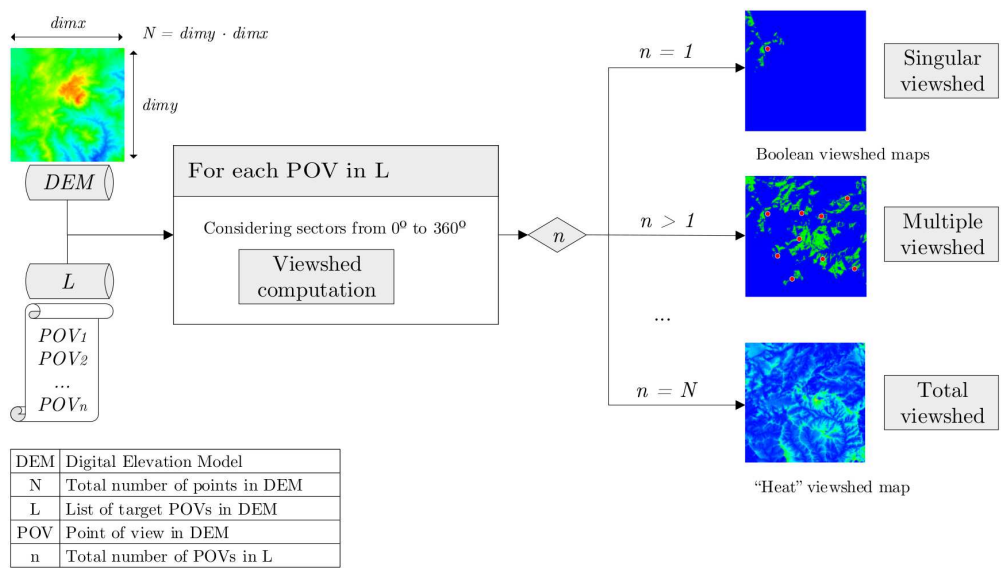}
\caption{Illustration of the different viewshed problems existing in the literature related to singular, multiple, and total viewsheds. This classification is based on the number of target points of view (POVs) in the Digital Elevation Model (DEM) chosen as input, each of which produces different outputs. Our work focuses on the total viewshed computation, which is the most computational demanding problem.}
\label{fig:1}
\end{figure}

The azimuthal sector discretization method is also included in some visibility modules from geographic information systems (GIS) applications such as ArcGIS~\citep{ARCINFO}, GRASS GIS~\citep{neteler2012grass}, and Google Earth. A very common tool provided by these programs is the viewshed computation, which is of great interest in many areas such as telecommunications, environmental planning, ecology, tourism, and archaeology~\citep{cauchi2015gpu,qarah2019fast,wang2019fast}. In these diverse fields, knowing the visibility in terrain is almost a requirement to achieve optimal results.

There are three types of viewshed problems in the literature (Figure~\ref{fig:1}) depending on the number of observers considered for calculation: (i) singular, (ii) multiple, and (iii) total viewsheds. The first is the simplest visibility problem which comprises the computation of the viewshed from one single observer at a certain height with respect to the ground. To address this problem, the DEM is usually divided into angular sectors in the plane around the POV. Then, different lines of sight (LoS), which correspond to the axis of every sector, start from the POV and are radially distributed towards the most distant areas. Every target point crossed over by this line is sequentially compared to the POV based on the elevation values in order to compute its visibility. The result is a boolean map containing visible and non-visible points from the POV in the terrain. Likewise, a multiple viewshed for several POVs can be obtained by repeating the above procedure for each POV and combining the viewshed results. However, a complete visibility analysis involves knowing the viewshed of every point in the terrain in all directions. This information can be used to address well-known problems such as siting multiple observers~\citep{cervillaICCS15}, and path planning with surveillance aims~\citep{li2010effective}. The first problem is related to finding the fewest possible number of POVs providing maximum viewshed for a certain area. The second involves designing a near optimal path aiming to achieve maximum terrain coverage. Both problems would be substantially simplified if the visibility of every point in the area is known beforehand~\citep{franklin1994higher}. This problem is known as total viewshed and is one of the most challenging visibility calculations due to its complexity and high computational cost. It involves obtaining the viewshed for each point in the DEM as a POV and then accumulating their visibility results as a new map where every point contains the viewshed value of the corresponding location measured, e.g., in $km^2$. Nowadays, most GIS software packages include specific modules for singular viewshed computation. Few provide multiple viewshed calculations, and when they do, it is carried out using task queues which demonstrate poor computational performance as they repeat the singular viewshed computation for each of the considered POVs.

In this work, we propose a new methodology called skewed Digital Elevation Model (sDEM) considering the total viewshed problem as a case study. It involves a complete restructuring of the DEM data in memory carried out prior to the computation of the total viewshed. The DEM is transformed into a new structure named skwDEM in which the data are aligned in memory to improve data locality in accessing the memory and, therefore, increasing the speed of processing. Through this approach, it is unnecessary to apply common techniques that reduce computational cost in total viewshed problems, such as considering a maximum visibility distance within a circular area around every POV. This methodology could also improve the performance of other applications that analyse relevant topographic features of the terrain surface such as slope and elevation.

The contributions of the paper are as follows:
\begin{itemize}
    \item We design a new methodology named sDEM (skewed Digital Elevation Model) for faster processing of terrain surface which substantially improves data locality in memory. In particular, this approach fully exploits the intrinsic parallelism of the total viewshed computation, achieving maximum performance through efficient memory access.
    \item We present different implementations for single-core, multi-core, single-GPU, and multi-GPU platforms. Each of these are compared with the state-of-the-art approaches.
\end{itemize}
The remainder of this paper is organized as follows: Section~\ref{work} presents the state-of-the-art in regards to the total viewshed computation. Section~\ref{background} reviews the background related to this research. Section~\ref{algorithm} explains the proposed sDEM methodology for computing the total viewshed and presents the implementation for multi-GPU systems. Section~\ref{experiments} compares the sDEM algorithm with the most commonly used GIS software and the state-of-the-art. Section~\ref{conclusions} discusses the results of this study.

\section{Related work}\label{work}
Terrain visibility, commonly known as viewshed analysis, is related to the problem of obtaining the area of the terrain visible from a given POV located at a certain elevation above the ground. This issue has been widely studied for many years given the mass of interpolation computations required to produce precise results~\citep{atallah1983dynamic,cabral1987bidirectional,fisher1992first,franklin1994higher,floriani1994visibility}. Authors usually use line of sight based algorithms such as R3, R2 or DDA~\citep{franklin1994geometric,kauvcivc2002comparison}. These methods project rays starting from the observer toward the boundary of the DEM to obtain the points included in processing. Another related strategy is XDraw~\citep{franklin1994geometric} which computes the LoS function in stages arranged as concentric squares centered on the position of the observer.

Many algorithms calculate the viewshed from a single POV, or from a small number of POVs at best. In~\citet{gao2011optimization} a singular viewshed implementation was developed for built-in GPU systems based on the LoS method and texture memory with bilinear interpolation. They achieve a speed-up up to 70x with respect to the sequential CPU implementation. The GPU implementation proposed by~\citet{stojanovic2013performance} achieves remarkable results in obtaining a boolean raster map instead of a map containing viewshed values. A novel reconfiguration of the XDraw algorithm for GPU context is described in~\citet{cauchi2015gpu} which outperforms CPU and GPU implementations of well-known viewshed analysis algorithms such as R3, R2, and XDraw. Furthermore, an efficient implementation of the R2 viewshed algorithm is carried out in~\citet{osterman2014io} with particular focus on input/output efficiency and obtaining significant results in contrast to the R3 and R2 sequential CPU implementations. The algorithm described in~\citet{zhao2013parallel} focuses on a two-level spatial domain decomposition method to speed-up data transfers and thus performs better than other well-known sequential algorithms. Other extended approaches are focused on obtaining the viewshed for multiple points~\citep{strnad2011parallel,song2016parallel}. More recent research is presented in~\citet{wang2019fast} where fast candidate viewpoints are obtained for multiple viewshed planning. These authors have also conducted a parallel XDraw analysis~\citep{dou2018fine,dou2019equal} to improve the results obtained by previous XDraw algorithms. 

Nevertheless, few studies address the total viewshed computation problem, and most of these focus on a simplified version. For example, the total viewshed in~\citet{dungan2018total} is obtained by drastically reducing the number of grid points to be processed. Likewise, the approach used in~\citet{brughmans2018introducing} computes the visibility of small areas and not for specific points. So far, the only algorithm that addresses the total viewshed problem on high resolution DEMs is the TVS algorithm proposed by~\citet{tabik2013simultaneous,tabik2014efficient}. It considers the closest points to the line of sight as a sample set of points stored in a structure called band of sight (BoS). In this approach, the distance to the axis determines the number of points in the BoS~\citep{floriani1994visibility}. Maximum memory utilization was achieved by reusing the points contained in the list and obtaining the viewshed for every aligned point in the particular sector. However, this algorithm has important limitations:
\begin{itemize}
    \item For a given POV, the analysis of the points inside the BoS is performed sequentially because it is impossible to know whether a target point is visible without knowing the state of the previous one. 
    \item The implementation of the data reuse of the BoS produces a significant overhead caused by the selection of the corresponding points for every direction. 
    \item It is not appropriate for implementation on high-throughput systems such as GPUs and Xeon-Phi architectures, because parallelism is limited to sector level. 
\end{itemize}
In this study, we propose computing total viewshed based on a compact and stable data structure with the aim of increasing data and computation reuse. Our proposal will be compared to the TVS algorithm~\citep{tabik2014efficient} and the most commonly used GIS software.

\section{Background: viewshed analysis}\label{background}
In this section, the basic concepts of the viewshed analysis are presented. Section~\ref{singularVS} describes the singular viewshed problem. Section~\ref{complexity} explains the complexity of the viewshed analysis. Section~\ref{tvsproblem} deals with the total viewshed problem.
\subsection{Singular viewshed}\label{singularVS}
As a starting point, most viewshed computation algorithms perform an azimuthal partition of the area. This division is carried out by splitting the area that surrounds the observer (POV) into $ns$ azimuthal sectors. Every sector is represented by its axis, and the closest points to this structure are usually considered for the viewshed analysis.

\begin{algorithm}[b!]
\small
\begin{algorithmic}
\STATE {{\bf point} $POV_{i,j,h} = DEM[i_0][j_0]$}
\STATE {$POV_h\mathrel{+}=h_0$}
\STATE {{\bf float} $VS=0$}
\FOR {$s=0,ns$}
\STATE {{\bf pointSet}  $axis = selectAxisPointSet(DEM,POV,s)$}
\STATE {$VS\mathrel{+}=linear\_viewshed(POV, axis, true)$  \, // forward }
\STATE {$VS\mathrel{+}=linear\_viewshed(POV, axis, false)$ // backward }
\ENDFOR
\STATE {$VS\mathrel{*}=(\pi / ns)$ \, // Papus theorem scaling}
\end{algorithmic}
\caption{$singular\_viewshed(DEM, i_0, j_0, h_0)$}
\label{alg:1}
\end{algorithm}

\begin{algorithm}[b!]
\small
\begin{algorithmic}
\STATE {{\bf global bool} $visible=true$}
\STATE {{\bf global float} $max\theta = -\infty$ \, // Max. angle}
\STATE {{\bf global pointSet} $visibleSet = \{\}$}
\STATE {{\bf do}}
\STATE {$ \quad \, {\bf point} \, \, T = axis.next()$}
\STATE {$ \quad \, point\_viewshed(POV,T)$}
\STATE {{\bf while} $T \mathrel{!}= axis.last()$}
\RETURN {$visibleSet.measure()$}
\end{algorithmic}
\caption{$linear\_viewshed(POV, axis, forward)$}
\label{alg:2}
\end{algorithm}

\begin{algorithm}[t!]
\small
\begin{algorithmic}
\STATE 	{{\bf float} $dist = \sqrt{(T_j-POV_j)^2+(T_i-POV_i)^2}$}
\STATE  {{\bf float} $\theta = (POV_h - T_h) \, / \, dist$}
\STATE  {{\bf bool} $prevVisible = visible$}
\STATE {{{\bf if} $(\theta>max\theta) \, visible = true$ {\bf else} $visible = false$}}
\STATE {{\bf bool} $startRS = \, !prevVisible ~ \& ~ visible$}
\STATE {{{\bf if} $(startRS) \, dist_0 = dist$}}
\STATE {{\bf bool} $endRS = prevVisible ~ \& ~ !visible$}
\STATE {{\bf if} $(endRS) \, visibleSet.add(dist_0,dist)$}
\end{algorithmic}
\caption{$point\_viewshed(POV, T)$}
\label{alg:3}
\end{algorithm}

Algorithm~\ref{alg:1} presents a general approach for the calculation of the viewshed considering one single $POV$ on a particular $DEM$ using a regular Cartesian grid. The initial coordinates of the observer's location are ($i_0$, $j_0$, $h_0$), where the last is the height of the observer measured from the eyes to the plane of the ground; the coordinates of the $POV$ structure used for the calculations are ($i$, $j$, $h$), where the last initially corresponds to the elevation of the point in the terrain; $VS$ is the structure that will accumulate the viewshed value; $axis$ is the set of points included in a particular sector $s$, and $selectAxisPointSet$ adds candidate points within the sector. This methodology analyzes the visibility in all the sectors based on the $linear\_viewshed$ function (Algorithm~\ref{alg:2}). It calculates the visible area for a certain sector with respect to an observer located on the axis of that sector. In practice, a linearized set of target points is considered as the visibility of every remaining point in $axis$ is computed following the direction from the nearest point to the furthest point.

\begin{figure}[b!]
    \centering
    \includegraphics[width=0.65\linewidth]{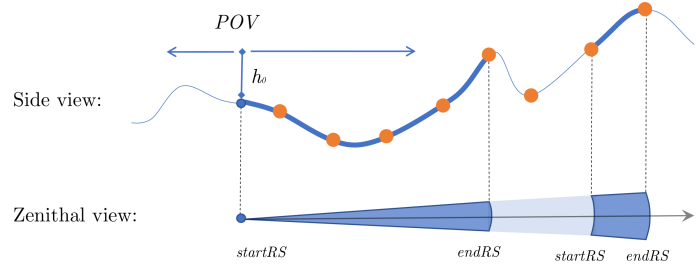}
    \caption{Side and zenithal views for a particular POV, with a specific height $h_0$, from which two segments are visible (represented both by blue thick segments). The corresponding visible ring-sectors are obtained for each one considering their starting points ($startRS$) and ending points ($endRS$).}
    \label{fig:2}
\end{figure}

As shown in Algorithm~\ref{alg:3}, a target point $T$ is visible from the $POV$ if its angular altitude $\theta$ is higher than all the previous ones considered in $axis$ ($max\theta$). Visible points on the axis are included in a set of points called $visibleSet$. In order to improve efficiency, only the starting and ending points of a segment are measured ($visibleSet.add$) and considered in processing. This methodology uses $startRS$ and $endRS$ variables to indicate whether a sequence of visible points has been found. First, the distance between the $POV$ and the first point found belonging to a visible section is measured in $dist_0$. Then, the final visible point of this visible segment is found and its distance with respect to the $POV$ is measured ($dist$). This process is repeated until all points on the axis are analyzed as shown in the side view in Figure~\ref{fig:2}. The projection of all visible segments throughout the sector results in the generation of visible sections, commonly known as ring-sectors. The area of every visible section ($A_{vs}$), considering sectors of one degree of opening, is computed as follows:
\begin{equation*}
    A_{vs} = (\pi/360)\cdot(R^2-r^2)
\end{equation*}
where $R$ and $r$ are the radius of the visible ring-sector related to the $endRS$ and $startRS$ values, respectively, with respect to a particular POV. Considering all the above, the viewshed for a location is the summation of the areas of all visible sections ($visibleSet.measure$). This approach reduces memory accesses and mathematical calculations as proven in \citet{tabik2014efficient}.

\subsection{Real problem complexity}\label{complexity}
The rotational sweep method significantly reduces the number of target points to analyze when computing the singular viewshed from $N$ to $s\cdot N^{1/2}$, where $N$ is the size of the DEM measured in points. Considering that a typical DEM greatly exceeds several millions of points and the discretization of the sector is rarely above the required accuracy, the accomplished reduction is between one and three orders of magnitude~\citep{stewart1998fast}. For example, the complexity to obtain the viewshed using point--to--point algorithms such as R3 is $O(N^{3/2})$, whereas it is reduced up to $O(s\cdot N^{1/2})$ using rotational sweep. However, there remains a large number of operations, which makes parallelism and supercomputing highly recommended for these sorts of approaches. In particular, one of the visibility problems that was considered unapproachable is the total viewshed computation, which is described in detail below.

\subsection{Total viewshed}\label{tvsproblem}
The problem of addressing the viewshed for all points in a particular area, represented by a DEM with $N$ points of observation, was almost impossible not long ago. The computation of the singular viewshed is very high demanding in computational terms and, therefore, repeating this procedure for every single point in the DEM would have been incredibly time-consuming on CPU. The inherent complexity of the problem is up to $O(N^3)$ if a non-optimized approach is applied $N$ times over a problem of $O(N^2)$ complexity. Nevertheless, using rotational sweep, the problem complexity can be reduced up to $O(s\cdot N^{3/2})$. Algorithm~\ref{alg:4} introduces the steps required to address the total viewshed problem, considering a DEM represented by a Cartesian grid with $dimy$ x $dimx$ points. The viewshed value, i.e, the visible terrain area for every point in the DEM is stored in the total viewshed matrix ($TVS$). This matrix has the same dimensions as the DEM and every cell is filled with a particular viewshed value obtained after performing the corresponding singular viewshed computation (previously shown in Algorithm~\ref{alg:1}). Some authors have observed that swapping the loops of Algorithm~\ref{alg:4} and Algorithm~\ref{alg:1} can significantly improve data locality in memory~\citep{stewart1998fast,tabik2014efficient}. This is one of the pillars on which our proposal is based.
\begin{algorithm}[htb!]
\small
\begin{algorithmic}
\FOR {$i=0,dimy$}
\FOR {$j=0,dimx$}
\STATE {$TVS[i][j] = singular\_viewshed(DEM, i, j, h_0)$}
\ENDFOR
\ENDFOR
\end{algorithmic}
\caption{$total\_viewshed(DEM, h_0)$}
\label{alg:4}
\end{algorithm}

\section{sDEM: a grid reorganization approach}\label{algorithm}
This section describes our proposed methodology called skewed Digital Elevation Model (sDEM) designed to improve data locality in memory for terrain surface analysis using the total viewshed computation problem as a case study. This approach takes into account the technical features of CPU (host) and GPU (device) processing units to take full advantage of the intrinsic parallelism of the total viewshed computation. For the sake of simplicity, in the rest of the paper we will refer to the input DEM as DEM and to the proposed modification of this structure as skwDEM.

\subsection{Proposed methodology}\label{methodology}
\begin{figure}[t!]
\centering
\subfigure[DEM; $2.5\,\sqrt{N}$ BoS]{
\resizebox*{0.319\textwidth}{!}{\label{fig:3-1}\includegraphics{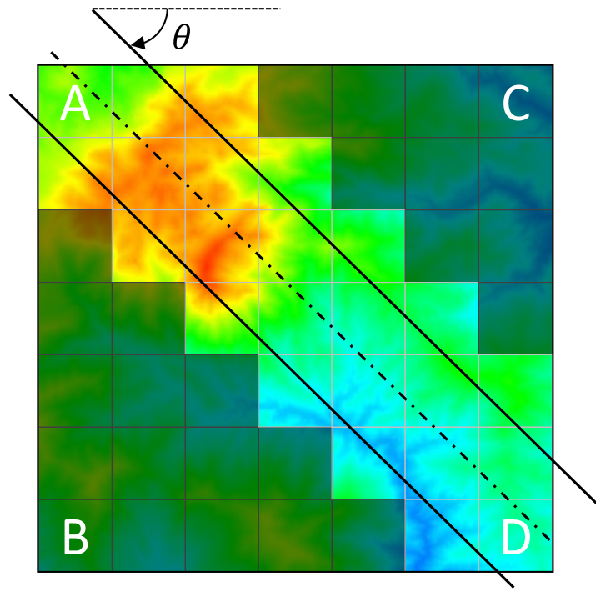}}}
\subfigure[DEM; $\sqrt{N}$ BoS]{
\resizebox*{0.319\textwidth}{!}{\label{fig:3-2}\includegraphics{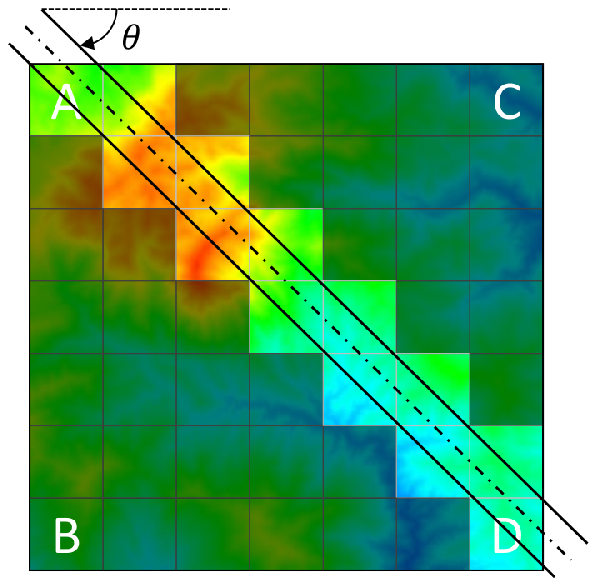}}}
\subfigure[skwDEM; $\sqrt{N}$ BoS]{
\resizebox*{0.319\textwidth}{!}{\label{fig:3-3}\includegraphics{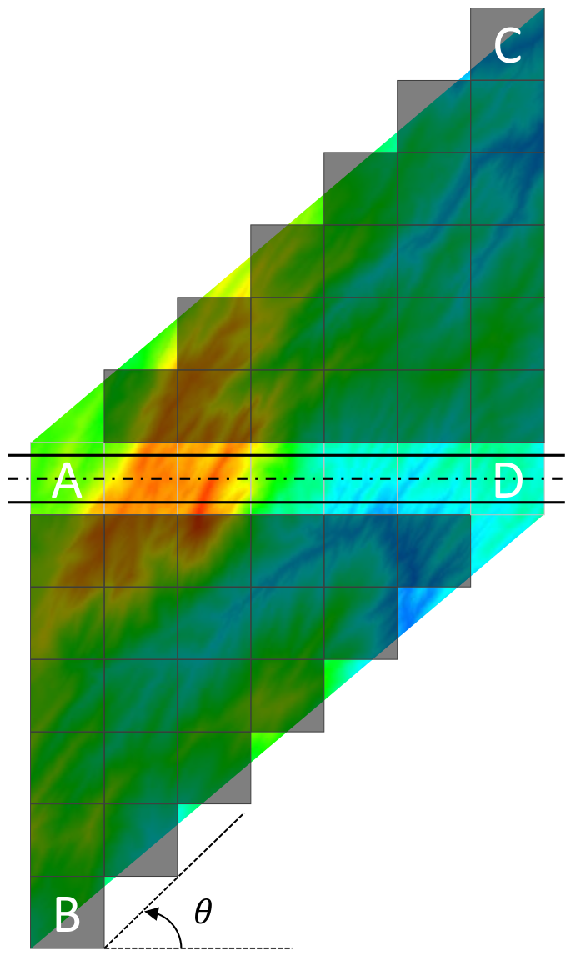}}}
\caption{Three examples of band of sight (BoS) on the plane, considering a sector $s=45^o$ for simplicity. The cells of the grid in dark color are not used for the visibility computation. (a) and (b) show two different BoS widths of $2.5\,\sqrt{N}$ and $\sqrt{N}$, respectively, with the same layout on the $dimy$ x $dimx$ DEM; whereas (c) presents the restructuring of the $2\cdot dimy$ x $dimx$ skwDEM considering a BoS width of $\sqrt{N}$. For the sake of clarity, A-D labels are located in the corner points of the DEM so that the restructuring approach can be visualized. Note that only one BoS is shown.}
\label{fig:3}
\end{figure}

Data reuse is key to the optimization of the total viewshed computation, so first we introduce the structure that will manage this process. This structure is called band of sight (BoS) and serves as the basis for the process of restructuring the DEM for every POV and sector. The BoS is used to find the closest points to the line of sight for any reference POV and given sector. Thus, choosing the right size for this structure is vital to improve data locality in accessing the memory. Figure~\ref{fig:3-1} and \ref{fig:3-2} show two BoS widths of $2.5\,\sqrt{N}$ and $\sqrt{N}$, respectively, considering a sector $s=45^o$ for the sake of simplicity; the cells of the grid in dark color are not considered for the visibility computation. The extensive statistical study conducted in~\citet{tabik2014efficient} proves that the size of this structure is not a determining factor, as long as it is of the order of $\sqrt{N}$. Therefore, our sDEM proposal uses the latter BoS size to address the data repetition problem.

Once the BoS width has been fixed, complete relocation of the data is performed from the DEM (Figure~\ref{fig:3-2}) to the skwDEM (Figure~\ref{fig:3-3}). This is a new DEM which is skewed in shape as a function of the BoS width. The use of this structure allows the exploitation of the existing parallelism without adversely affecting the precision of the results based on the following considerations:
\begin{itemize}
    \item We apply the Stewart sweep method \citep{stewart1998fast} which states that an outer loop iterates over the sectors and an inner loop over the points in the DEM. It is the only model that guarantees the reuse of data aligned in every direction.
    \item Given a sector, all the possible parallel bands of sight that cross the DEM are built simultaneously. We apply the interpolation method based on a simplified version of Bresenham's algorithm, which is commonly used for line rasterization. This algorithm was chosen for its high speed as well as maintaining sufficient fidelity to the problem under consideration.
    \item For each sector, the relocation is applied only once to the entire DEM. For example, in the particular case of considering 180 sectors, the data relocation takes place 180 times and always before starting the viewshed computation. Thus, the relocation only depends on the selected sector. Another advantage is that this method is especially appropriate for processing on the GPU as it aims to reduce the conditional structures to the maximum, hence avoiding the well-known thread divergence penalty.
\end{itemize}

\begin{figure}[t!]
\centering
\subfigure[DEM]{
\resizebox*{.22\textwidth}{!}{\label{fig:4-1}\includegraphics{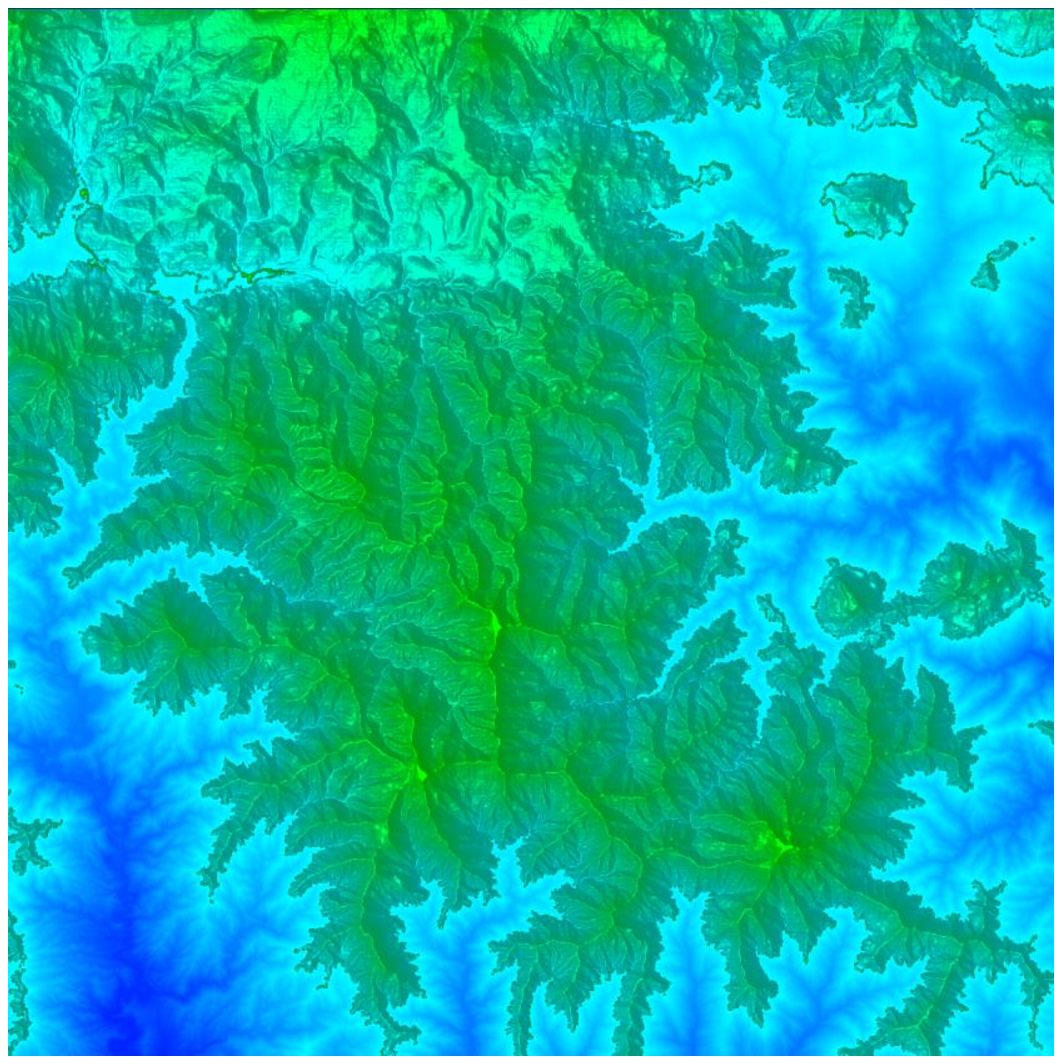}}}\hspace{4pt}
\subfigure[Reorganized skwDEM]{
\resizebox*{.23\textwidth}{!}{\label{fig:4-2}\includegraphics{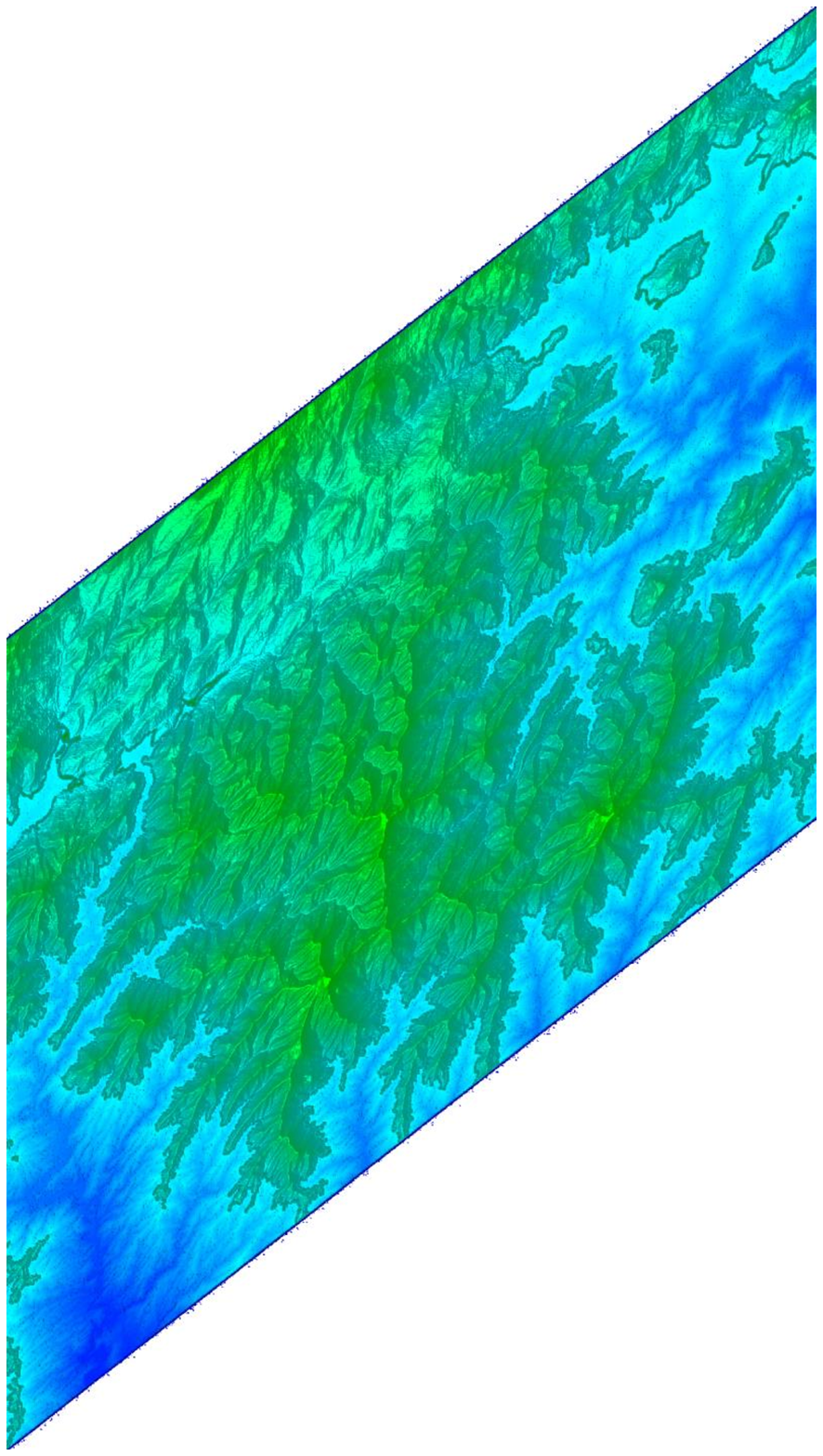}}}\hspace{4pt}
\subfigure[Compacted$_1$ skwDEM]{
\resizebox*{.23\textwidth}{!}{\label{fig:4-3}\includegraphics{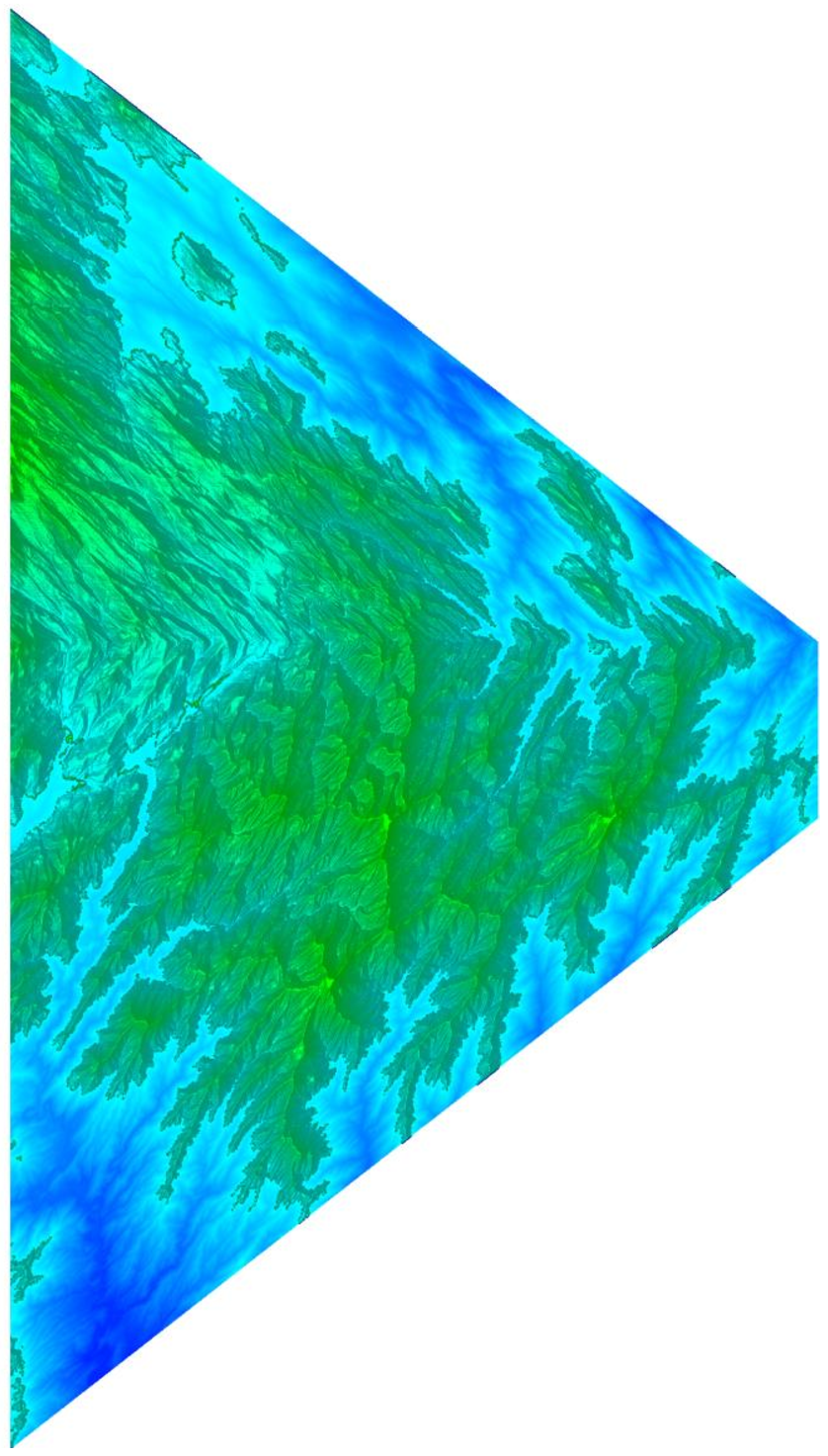}}}\hspace{4pt}
\subfigure[Compacted$_2$ skwDEM]{
\resizebox*{.23\textwidth}{!}{\label{fig:4-4}\includegraphics{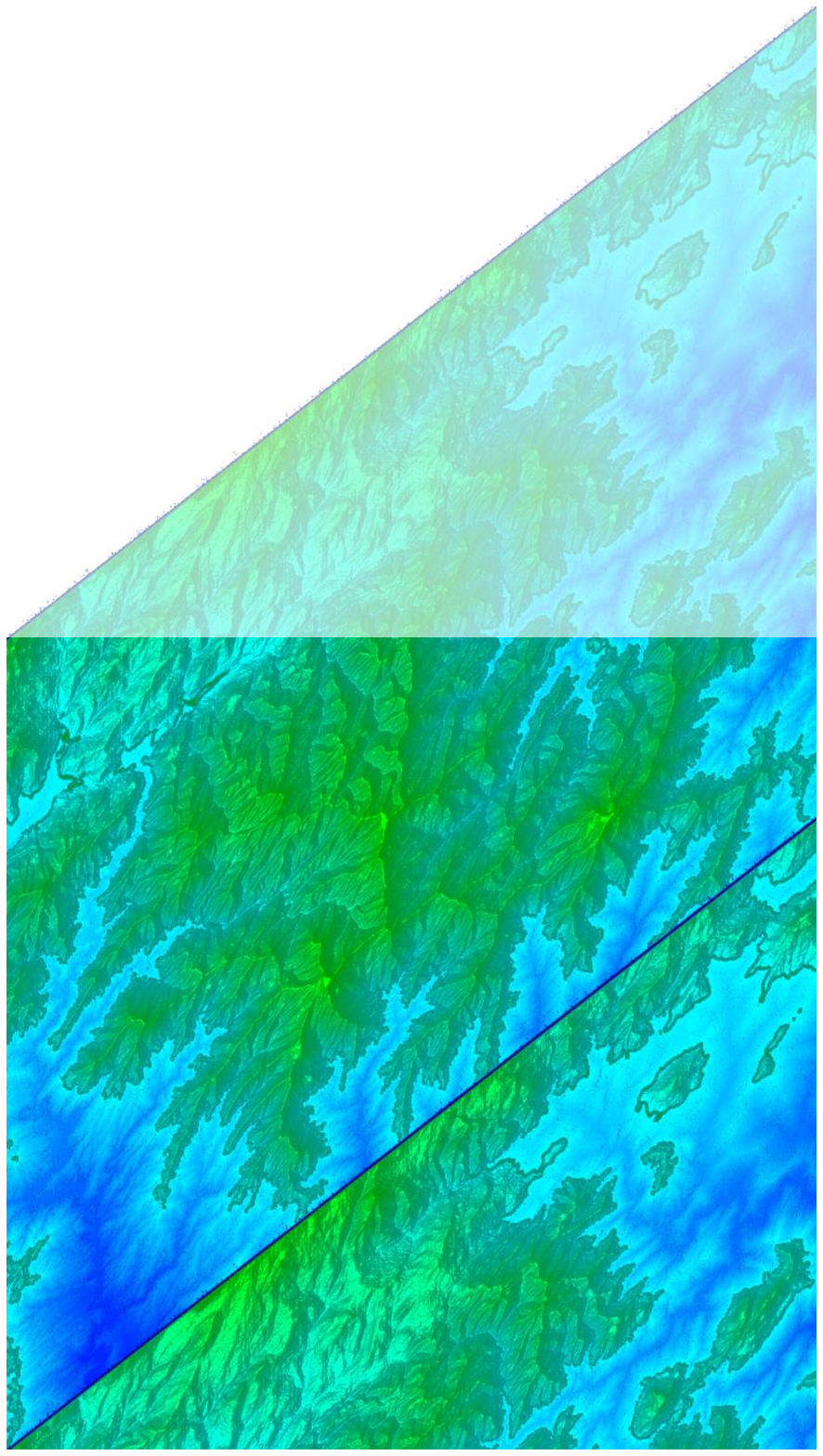}}}
\caption{The DEM and three possible results when applying our array redistribution procedure considering sector $s=45^o$, for the sake of simplicity. (a) presents the input DEM, (b) shows the skwDEM used in this work; (c) and (d) introduce two possible ways of compacting the data.} 
\label{fig:4}
\end{figure}

Figure~\ref{fig:4} shows the different possible redistributions of rows and columns using the DEM of the Montes de Malaga Natural Park (Malaga, Spain), complementing Figure~\ref{fig:3}. Data of the same latitudes are stored contiguously in memory in the DEM (Figure~\ref{fig:4-1}); that is, the outer loop iterates from north to south, whereas the inner loop iterates from west to east. Using the interpolation method, all parallel segments from the DEM (Figure~\ref{fig:4-1}) are projected into the skwDEM structure (Figure~\ref{fig:4-2}) so that the number of non-null elements of both structures matches. In this reorganized dataset, unlike the original, all the points in a given sector are placed in the same row and, therefore, memory accesses are sequentially performed increasing locality.

The reorganized matrix shown in Figure~\ref{fig:4-2} could later be compacted by aligning all data to the left of the structure (Figure~\ref{fig:4-3}), or relocating the data within the upper light color triangle to the lower right area of the structure, thus forming a $dimy$ x $dimx$ square structure (Figure~\ref{fig:4-4}). This second method aims to further compact the information to make memory access as regular as possible at the cost of an increase in complexity, and hence building time. Although both approaches seem to fit better for GPU processing in theory, they have not revealed significant differences in practice. Therefore, only the simplest and fastest approach shown in Figure~\ref{fig:4-2} is used for building the skwDEM structure in all implementations of the sDEM algorithm.

\begin{figure}[b!]
\centering
\subfigure[Main loop]{
\resizebox*{0.487\textwidth}{!}{\label{fig:5-1}\includegraphics{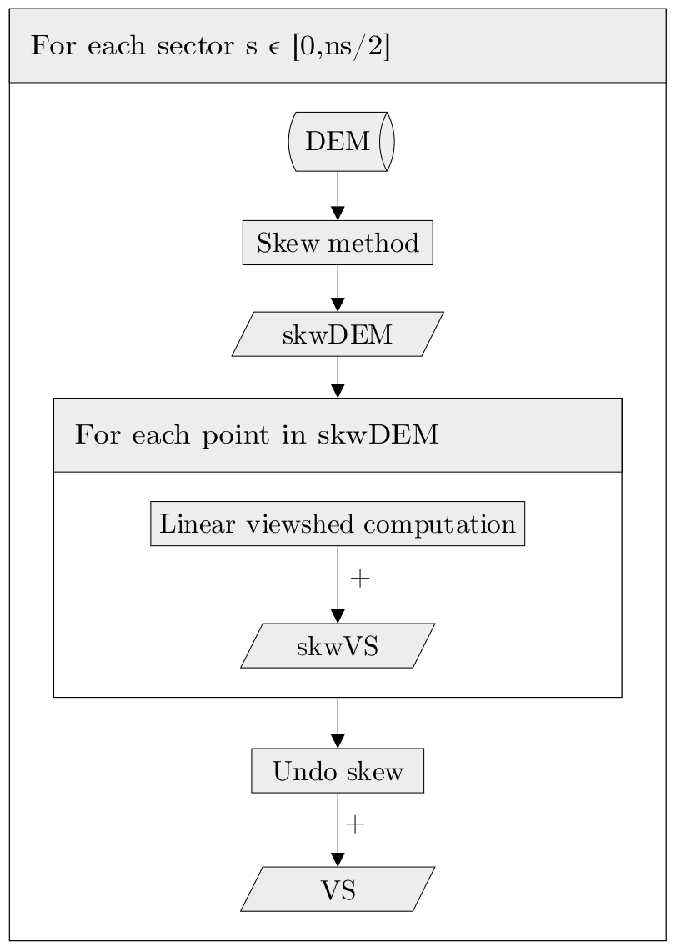}}}
\subfigure[Single iteration]{
\resizebox*{0.487\textwidth}{!}{\label{fig:5-2}\includegraphics{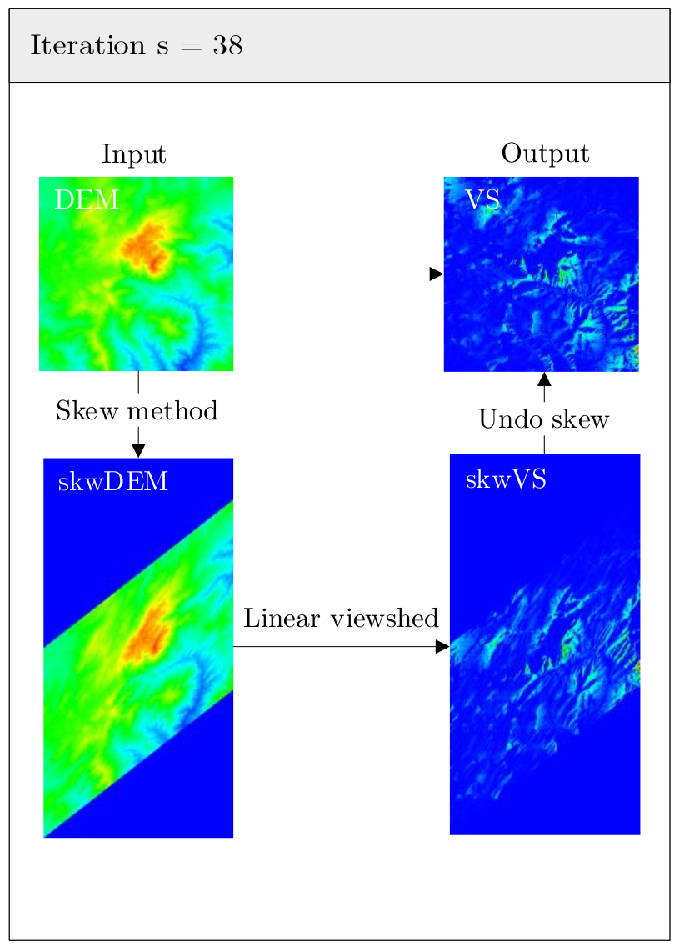}}}
\caption{Flowchart of our sDEM proposal for total viewshed computation showing (a) the steps inside the main loop which runs through all the sectors and (b) the outputs obtained in a single iteration of the same loop. In the last, null and low values are represented in blue; whereas red cells represent maximum values. The mathematical symbol (+) represents the process of accumulation.}
\label{fig:5}
\end{figure}

Given the above, our sDEM methodology can be described as follows (Figure~\ref{fig:5}):
\begin{enumerate}
    \item For each sector $s \in [0,ns/2]$, do:
    \begin{enumerate}
        \item Create the $2\cdot dimy$ x $dimx$ skwDEM, which is unique to each sector, from the $dimy$ x $dimx$ DEM and $s$. 
        \label{das}
        \item Calculate the horizontal $(A,D)$ and vertical $(B,C)$ limits of the skwDEM structure, which depend on the sector $s$.
        \item Let POV$_{i,j}$ be the point of view with $i$ and $j$ coordinates. For each point POV$_{i,j} \in$ skwDEM with $i\in[A,D]$ and $j\in[B,C]$, do:
        \begin{enumerate}
            \item Compute the linear viewshed considering sectors $s$ and $s+180^o$, i.e., analyze to the right and to the left the points in the row to which POV$_{i,j}$ belongs in the skwDEM.
            \item Accumulate the viewshed result in a new structure called skwVS, which is similar in size to the skwDEM structure.
        \end{enumerate}
        \item Transform skwVS into the VS structure (viewshed on the DEM) by undoing the operations performed in Step~\ref{das}. This procedure includes Pappus's theorem and also corrects the deformation introduced by the skwDEM. Accumulate the results in VS.
    \end{enumerate}
\end{enumerate}
The viewshed computation of all sectors is an embarrassingly parallel task because the computation of each sector is independent. This reduces the total viewshed problem to the calculation of $ns/2$ times the singular viewshed problem for every point in the skwDEM. Moreover, by skewing before carrying out the viewshed computation for a given sector, sDEM ensures that each BoS needed by each point as a POV has been previously built and included in the skwDEM. Following this approach, each row of the skwDEM corresponds to the static BoS of every point included in it for a given sector; that is, if a point is a POV, the remainder of the points in its row will form the BoS for that point and sector. As the skwDEM depends on the sector, it has to be reconstructed only $ns/2$ times. In contrast, the BoS structure used in~\citet{tabik2014efficient} must be reconstructed for each point and sector, which corresponds to $N\cdot ns$ times.

\subsection{Multi-GPU implementation}\label{implementation}
Since this particular problem is similar to matrix processing, our proposal to accelerate the calculation of the total viewshed focuses on exploiting the intrinsic parallelism of this procedure through the use of GPU processing units. In practice, the $ns/2$ sectors are distributed among all available devices so that each one is in charge of processing a similar number of given sectors, which will depend on the chosen scheduling. Every device sequentially launches three kernels, which will be further described, in order to process the viewshed of all points in the DEM for the corresponding sectors. In this way, each device contains partial viewshed results which are added up by the host in a final stage to obtain the total viewshed. Our method is used to avoid dependencies between threads while performing the viewshed computation. In addition, we will denote block and thread identification numbers as $b_{id}$ and $t_{id}$, respectively, and thread block dimension as $b_{dim}$.

\begin{algorithm}[b!]
\small
\begin{algorithmic}
\STATE {{\bf int} $i = b_{idy} \cdot b_{dim} + t_{idy}$}
\STATE {{\bf int} $j = b_{idx} \cdot b_{dim} + t_{idx}$}
\STATE {{\bf float} $y = tan(s) \cdot j$}
\STATE {{\bf int} $dest = y$}
\STATE {{\bf float} $r = y - dest$}
\STATE {{\bf int} $p = dimy + i - dest$}
\IF {$(i<dimy)\,\&\,(j<dimx)$}
\STATE {$skwDEM[p][j]\mathrel{+}=(1 - r) \cdot DEM[i][j]$}
\STATE {$skwDEM[p-1][j]\mathrel{+}=r \cdot DEM[i][j]$}
\ENDIF
\end{algorithmic}
\caption{Kernel-1 in charge of generating the $skwDEM$ structure from the $DEM$ ($0^o\leq s\leq45^o$)}
\label{alg:6}
\end{algorithm}

\subsubsection{Kernel-1: obtaining the skwDEM structure}
This kernel is in charge of transforming the $DEM$ into the $skwDEM$ structure for a given sector (Algorithm~\ref{alg:6}). In this new model and for the chosen direction, points located consecutively in the terrain are also stored sequentially in memory, which improves the performance of the memory accesses. This is achieved by using the interpolation based on Bresenham's algorithm to soften the projection of the points. Every thread is in charge of interpolating the corresponding point according to its 2D thread identification number defined by $i$ and $j$ variables. 

Regarding the implementation, this kernel is launched using $C_{by}=dimy/8$ and $C_{bx}=dimx/8$ 2D threads blocks with 8 threads per block, so as not to exceed the maximum register file size shared between thread blocks, thus avoiding scheduling problems.

\subsubsection{Kernel-2: viewshed computation on the skwDEM}
This kernel computes the viewshed for every point in the $skwDEM$ and a given sector, obtaining as a result the $skwVS$ matrix. The pseudo-codes of this kernel are shown in Algorithms~\ref{alg:7} and \ref{alg:8}, where each thread manages a particular point $POV_t\in skwDEM$, where $t=\{i,j\}$ is the corresponding two dimensional thread. The variable $h$ contains both the observer's height and the elevation of the location in this case. Each thread obtains its computation range of non-zero values from the corresponding row in the $skwDEM$, which is contained in the $nzSet$ structure. Then, each thread computes its visibility forward and backward across the row to which it belongs in a process called linear viewshed computation. The resulting viewshed value is thereafter stored in its corresponding position of the $skwVS$ matrix. 

This kernel is launched with $2\cdot C_{by}$ and $C_{bx}$ 2D threads blocks using twice as many thread blocks as in the y-dimension according to the size of the $skwDEM$ matrix.

\begin{algorithm}[h!]
\small
\begin{algorithmic}
\STATE {{\bf int} $i = b_{idy} \cdot b_{dim} + t_{idy}$}
\STATE {{\bf int} $j = b_{idx} \cdot b_{dim} + t_{idx}$}
\STATE {{\bf float} $r = (1.0 / cos(s))^2$}
\STATE {{\bf float} $cv=0$}
\IF {$(i<2\cdot dimy)\,\&\,(j<dimx)$}
\STATE {{\bf float} $h = skwDEM[i][j] + h_0$}
\STATE {$cv\mathrel{+}=linear\_viewshed(i,j,h,skwDEM,true)$}
\STATE {$cv\mathrel{+}=linear\_viewshed(i,j,h,skwDEM,false)$}
\STATE {$skwVS[i][j] = cv \cdot r$}
\ENDIF
\end{algorithmic}
\caption{Kernel-2 in charge of the $skwVS$ computation on the skwDEM}
\label{alg:7}
\end{algorithm}
\begin{algorithm}[h!]
\small
\begin{algorithmic}
\STATE {{\bf int} $dir$}
\STATE {{\bf if} $\,(forward)$ {\bf then} $dir=1$  {\bf else} $dir = -1$}
\STATE {{\bf int} $k = j + dir$}
\STATE {{\bf bool} $visible,above,opening,closing$}
\STATE {{\bf float} $open\Delta d, \Delta d, \theta, max\theta = -\infty$}
\WHILE 	{$k\in nzSet$}
\STATE {$\Delta d = |k - j|$}
\STATE {$\theta = (skwDEM[i][k] - h) / \Delta d$}
\STATE {{{\bf if} $(\theta>max\theta)$ {\bf then} $above=true$}}
\STATE {$opening = above ~ \& ~ !visible$}
\STATE {$closing = \, !above ~ \& ~ visible$}
\STATE {$visible = above$}
\STATE {$max\theta = max(\theta,max\theta)$}
\STATE {{\bf if} $\,(opening)$ {\bf then} $\,open\Delta d = \Delta d$}
\STATE {{\bf if} $\,(closing)$ {\bf then} $\,cv\mathrel{+}=\Delta d\cdot\Delta d - open\Delta d\cdot open\Delta d$}
\STATE {$k \mathrel{+}=dir$}
\ENDWHILE
\RETURN{cv}
\end{algorithmic}
\caption{$linear\_viewshed(i,j,h,skwDEM,forward)$}
\label{alg:8}
\end{algorithm}

\subsubsection{Kernel-3: obtaining the final viewshed on the DEM}
Once the viewshed is computed on the $skwDEM$ for every POV and stored in the $skwVS$ matrix, this kernel transforms the latter structure by undoing the rotation performed in Kernel-1 to obtain the final viewshed $VS$ matrix on the DEM. The pseudo-code in charge of performing this procedure is presented in Algorithm~\ref{alg:9}. This kernel is also launched with the same configuration as Kernel-1.

\begin{algorithm}[hbt!]
\small
\begin{algorithmic}
\STATE {{\bf int} $i = b_{idy} \cdot b_{dim} + t_{idy}$}
\STATE {{\bf int} $j = b_{idx} \cdot b_{dim} + t_{idx}$}
\STATE {{\bf float} $y = tan(s) \cdot j$}
\STATE {{\bf int} $dest = y$}
\STATE {{\bf float} $r = y - dest$}
\STATE {{\bf int} $p = dimy + i - dest$}
\IF {$(i<dimy) ~ \& ~ (j<dimx)$}
\STATE {{\bf float} $skwVS_a = skwVS[p][j]$}
\STATE {{\bf float} $skwVS_b = skwVS[p-1][j]$}
\STATE {$VS[i][j]\mathrel{+}=(1 - r) \cdot skwVS_a + r \cdot skwVS_b$}
\ENDIF
\end{algorithmic}
\caption{Kernel-3 in charge of transforming the $skwVS$ on the skwDEM to the $VS$ structure on the DEM ($0^o\leq s\leq45^o$)}
\label{alg:9}
\end{algorithm}

\subsubsection{Scheduling multi-GPU processing on the host}
In order to perform the total viewshed calculation in a multi-GPU system, several steps must be followed as shown in Algorithm~\ref{alg:10}. First, we must reserve the required memory spaces to allocate the different structures in all the available devices. Then, the $DEM$ can be transferred from the host to each device ($dev$) of the total available devices ($nd$). The target number of sectors $ns/2$ will be distributed among the different devices so that the workload is balanced. Each device will execute the three above-mentioned kernels accumulating the result of the viewshed computation in their private $VS_d$ structure, considering all the points and every target sector. Finally, these structures are transferred from the devices to the host so that a final parallel reduction can be performed, obtaining the total viewshed $VS$ final result.

\begin{algorithm}[htb!]
\small
\begin{algorithmic}
\FOR {$d=0,nd$}
\STATE {$dev_d \leftarrow Allocate(|DEM|,|skwDEM|,|skwVS|,|VS_d|)$}
\ENDFOR
\FOR {$d=0,nd$}
\STATE {$dev_d \leftarrow MemcpyAsyncH2D(DEM)$}
\ENDFOR
\FOR {$s=0,ns/2$}
\STATE {{\bf int} $d = s \: \% \: nd$}
\STATE {$dev_d \leftarrow Kernel-1,2,3\,(s)$}
\STATE {$dev_d \leftarrow MemcpyAsyncD2H(VS_d)$}
\ENDFOR
\FOR {$d=0,nd$ {\bf parallel}}
\STATE {$VS\mathrel{+}=VS_d$}
\ENDFOR
\end{algorithmic}
\caption{Host code in charge of scheduling the work for the different devices}
\label{alg:10}
\end{algorithm}

\section{Experiments}\label{experiments}
This section assesses the performance of our sDEM proposal with respect to well-known GIS software and the state-of-the-art. Section~\ref{setup} explains the experimental setup. Section~\ref{comparison} presents the comparison between sDEM and GIS software in addressing the multiple viewshed problem. Section~\ref{analysis} evaluates the computational performance of sDEM compared to the state-of-the-art algorithm using the total viewshed problem as a case study in three scenarios: (i) sector viewshed computation for a random direction, (ii) average sector viewshed computation, and (iii) total viewshed map generation.

\subsection{Experimental setup}\label{setup}
We select two operating systems (OSs) for the experiments according to their requirements:
\begin{itemize}
\item Windows OS: Windows 10 with an Intel(R) Core(TM) i5-6500 CPU @3.20GHz with 4 cores (4 threads) and 8GB DDR4 RAM.
\item Linux OS: Ubuntu 16.04.5 LTS with an Intel(R) Xeon(R) E5-2698 v3 @2.30GHz with 16 cores (32 threads) and 256GB DDR4 RAM, along with four GTX 980 Maxwell GPUs with 2048 CUDA cores, 16 SMs, 1.12GHz, and 4GB GDDR5 each one.
\end{itemize}
The experiment presented in Section~\ref{comparison} is executed on Windows OS because GIS software is usually developed for this specific operating system; whereas the experiment described in Section~\ref{analysis} is executed on Linux OS to obtain an optimal measurement of the computational performance. These experiments were designed to be as representative as possible of a real problem where obtaining the visibility in a particular direction or region is necessary. This is the reason why three DEMs of the Montes de Malaga Natural Park (Malaga, Spain) were considered. Each has 10 meters resolution and different extents (Table~\ref{tab:2}). The calculation of the total viewshed is not only limited to this area of interest, but includes the surrounding region which contains both flat and mountainous areas. Observers are considered to be located 1.5 meters above the ground. 

\begin{table}[b!]
\tbl{Different DEMs used in the experiments.}
{\begin{tabular}{lcccc} \toprule
 & & \multicolumn{2}{l}{UTM} \\ \cmidrule{3-5}
 Dataset & Dimension & Zone\textsuperscript{a} & Easting & Northing \\ \midrule
 DEM10m-2k & 2000x2000 & 30S & 0310000mE & 4070000mN \\
 DEM10m-4k & 4000x4000 & 30S & 0360000mE & 4100000mN \\
 DEM10m-8k & 8000x8000 & 30S & 0360000mE & 4140000mN \\ \bottomrule
\end{tabular}}
\tabnote{\textsuperscript{a}Latitude band designator.}
\label{tab:2}
\end{table}

Regarding the implementation, the OpenMP API is used to enable the multi-threaded execution of every selected sector with dynamic scheduling since it has proved to obtain the best performance. Host codes are compiled using the g\texttt{++} 5.4 open-source compiler with ffast-math, fopenmp, and O2 optimization flags. CUDA files make use of the NVIDIA NVCC compiler from the CUDA compilation tools V10.0.130. The multi-core implementation of our proposal is launched with the maximum number of threads available in the system, the same way that single-GPU and multi-GPU implementations are configured to operate the devices at full capacity.

The GIS software used for comparison includes ArcGIS 10.7, specifically the Spatial Analyst extension which contains the Viewshed 2 (VS-2), Viewshed (VS), and Visibility (VI) tools. Google Earth Pro 7.3 and QGIS-GRASS 3.10.2 are also used.

\subsection{Comparison with GIS software}\label{comparison}
\begin{figure}[h!]
    \centering
    \includegraphics[width=\columnwidth]{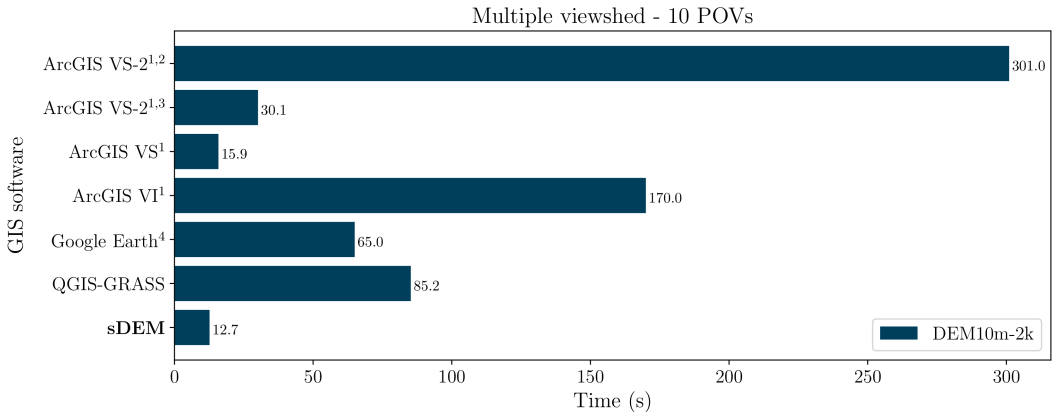}
    \caption{Computational performance comparison between the sDEM proposal and commonly used GIS software in solving the multiple viewshed problem, considering 10 POVs randomly located in the DEM10m-2k. Single-thread execution is considered to obtain the run-time of those programs not otherwise indicated. VS, VS-2, and VI are the ArcGIS Spatial Analyst tools Viewshed, Viewshed 2 and Visibility, respectively. QGIS-GRASS uses the r.viewshed module. $^1$multi-thread execution was required using the maximum number of cores available. $^2$using PERIMETER\_SIGHTLINES parameter. $^3$using ALL\_SIGHTLINES parameter. $^4$Google Earth does not have multiple/total viewshed computation capability so the average time in computing singular viewshed has been multiplied by the number of POVs considered.}
    \label{fig:6}
\end{figure}

A fair comparison in the context of this work would be to compare the total viewshed computation using our approach and other GIS software/tools. However, as there does not exist any public software/tool to compute total viewshed, we will compare the results only for few points. This experiment assesses the computational performance of sDEM and the most used GIS software in solving the multiple viewshed problem. In particular, the objective is to compute the accumulated viewshed considering 10 POVs located randomly in the DEM10m-2k using a random number generator to determine the coordinates. Single-threaded implementations were used for our sDEM proposal, QGIS-GRASS, and Google Earth; whereas ArcGIS has to be executed using all available cores. Moreover, the execution time obtained using Google Earth results from extrapolating the singular viewshed computation to the current case of 10 POVs since this software does not support this operation. Figure~\ref{fig:6} shows the time each software requires to complete the multiple viewshed task. Our sDEM proposal outperforms every analyzed GIS software: ArcGIS VS-2 with two different configurations (23.7x, 2.4x), ArcGIS VS (1.3x), ArcGIS VI (13.4x), Google Earth (5.1x), and QGIS-GRASS (6.7x). Although sDEM achieves significant results in this multiple viewshed computation, the greatest gain is given in the total viewshed computation discussed below.

\subsection{Total viewshed analysis}\label{analysis}
To the best of our knowledge, our sDEM proposal and the TVS algorithm~\citep{tabik2014efficient} are the only approaches in the literature capable of performing the total viewshed computation on entire datasets without carrying out prior reductions in workload. In order to achieve a fair analysis, a size of $dimx$ has been chosen for the BoS in the case of the TVS algorithm so that this structure coincides with the number of points processed per row in the sDEM algorithm. Thus, the workload is similar for both TVS and sDEM algorithms making it possible to perform a computational performance comparison using speed-up and throughput values. We implemented single-threaded, multi-threaded, single-GPU, and multi-GPU versions of our sDEM proposal and compared them to the single-thread implementation of the TVS algorithm. Three experiments were used to assess the performance of the total viewshed computation.

\subsubsection{Sector viewshed considering a random sector}
\begin{figure}[h!]
\centering
\subfigure[Speed-up curves]{
\resizebox*{0.487\textwidth}{!}{\label{fig:7-1}\includegraphics{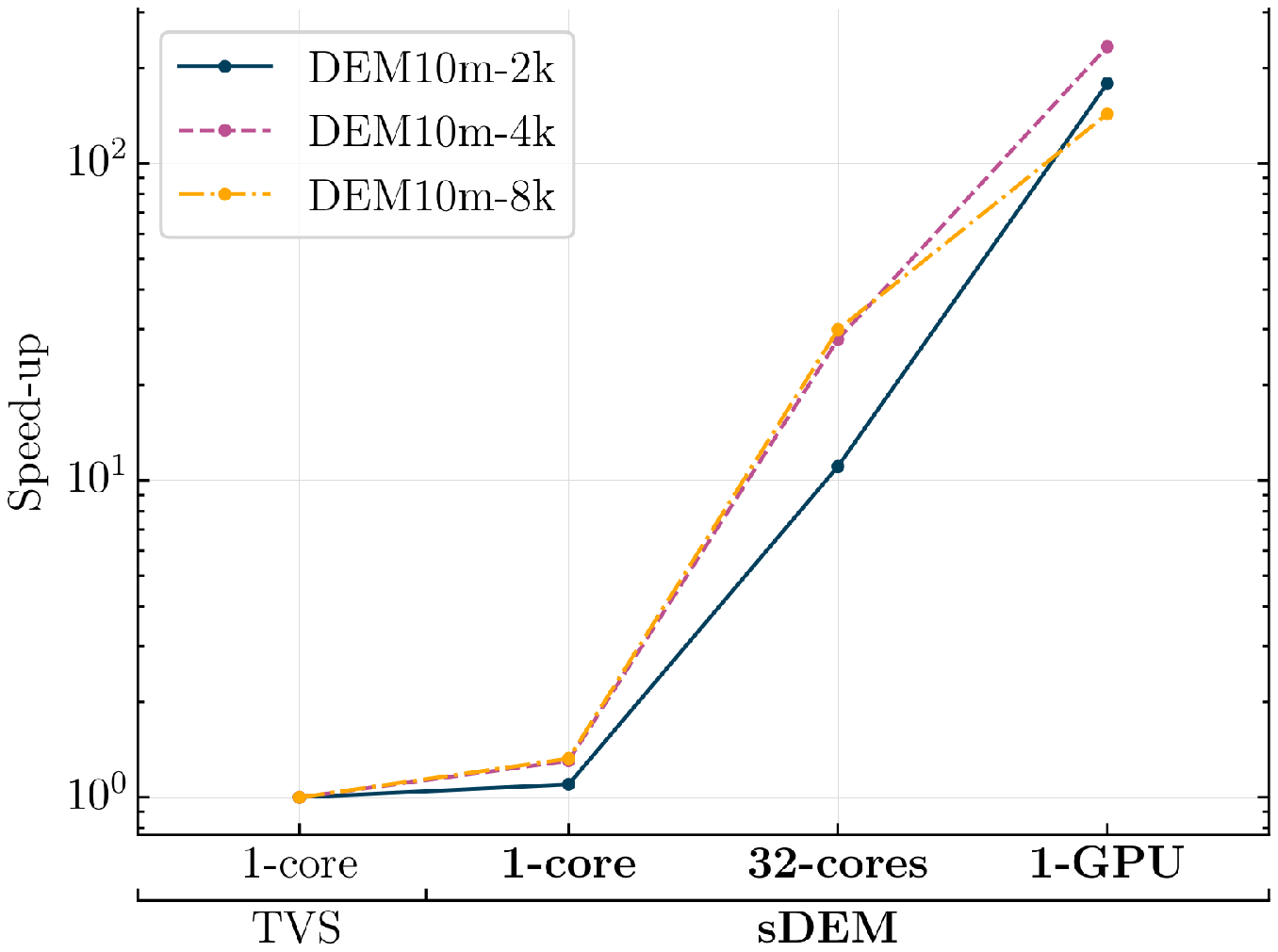}}}
\subfigure[Diagram of throughputs]{
\resizebox*{0.487\textwidth}{!}{\label{fig:7-2}\includegraphics{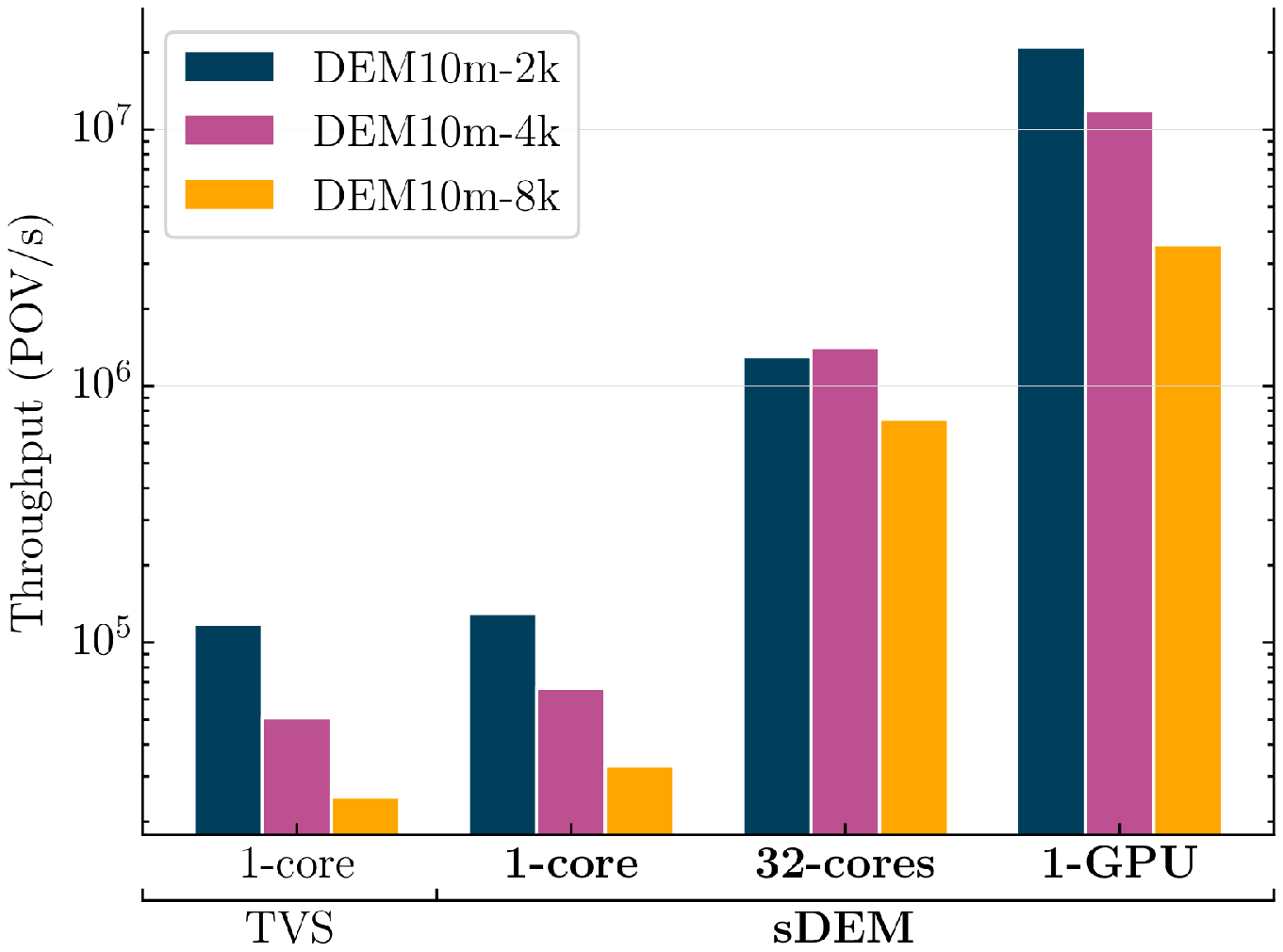}}}
\caption{Speed-up curves and throughput diagrams for the state-of-the-art total viewshed algorithm, TVS~\citep{tabik2014efficient} and the different implementations of our sDEM proposal using single-core, multi-core, single-GPU, and multi-GPU platforms to compute singular viewshed for a randomly selected sector, $s=10^o$ (BoS size of \textit{dimx} points for the TVS algorithm). Each color corresponds to a particular dataset. Logarithmic scale is used.}
\label{fig:7}
\end{figure}

The computational performance of our sDEM proposal was analyzed and compared with the TVS algorithm in computing total viewshed considering a single random sector, where the sector $10^o$ is selected. Figure~\ref{fig:7} presents the acceleration curves and the throughput results (POVs processed per second) using three DEMs. Our algorithm outperforms the TVS algorithm, in the best case achieving a maximum acceleration up to 233.5x using the 1-GPU implementation on DEM10m-4k. Throughput results show that this variable increases approximately 178.9x for the same implementation on DEM10m-2k. Multi-GPU implementations are not considered due to the low workload when distributing one sector across more than one device.

\subsubsection{Sector viewshed based on average values}
\begin{figure}[t!]
\centering
\subfigure[Speed-up curves]{
\resizebox*{0.487\textwidth}{!}{\label{fig:8-1}\includegraphics{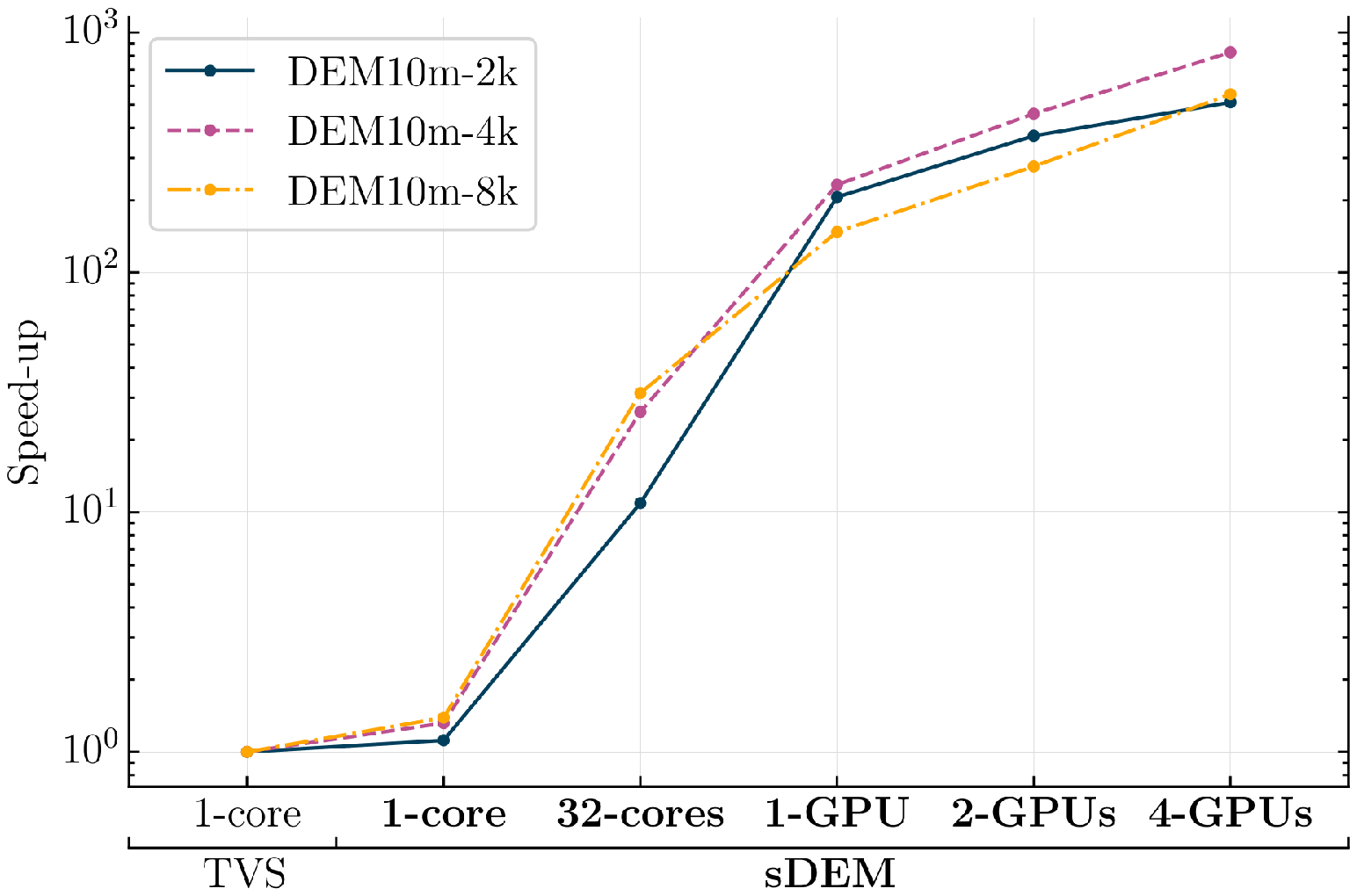}}}
\subfigure[Diagram of throughputs]{
\resizebox*{0.487\textwidth}{!}{\label{fig:8-2}\includegraphics{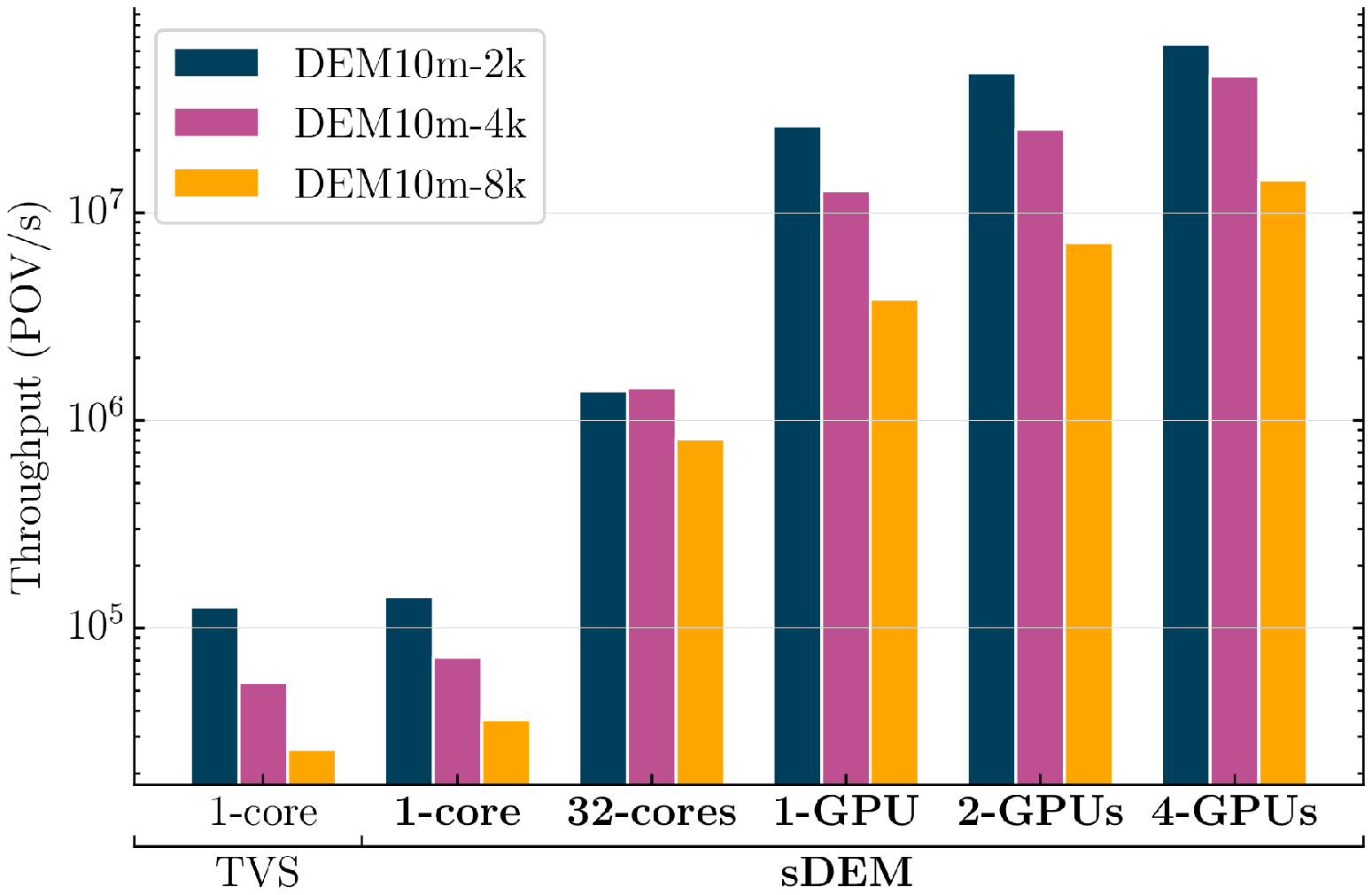}}}
\caption{Speed-up curves and throughput diagrams for the state-of-the-art total viewshed algorithm, TVS~\citep{tabik2014efficient} and the different implementations of our sDEM proposal using single-core, multi-core, single-GPU, and multi-GPU platforms to compute sector viewshed. Directions fulfilling $0^o<s<45^o$ are considered to obtain average values per sector (BoS size of \textit{dimx} points for the TVS algorithm). Each color corresponds to a particular dataset. Logarithmic scale is used.}
\label{fig:8}
\end{figure}

In this experiment, unlike the prior one, the direction range is selected from $0^o$ up to $45^o$ to obtain average values per sector. This choice of design lies in the fact that single-threaded executions of TVS and sDEM required several weeks to complete when using a higher range. Moreover, results within this range are representative and can be extrapolated to any target range. Figure~\ref{fig:8} introduces the acceleration curves and the throughput results achieved. Best-studied cases show that the maximum speed-up result achieved is up to 827.3x with the 4-GPUs implementation with respect to the TVS algorithm considering DEM10m-4k. Throughput results show that this variable increases approximately 511.8x for the same implementation on DEM10m-2k.

\begin{figure}[b!]
    \centering
    \includegraphics[width=\columnwidth]{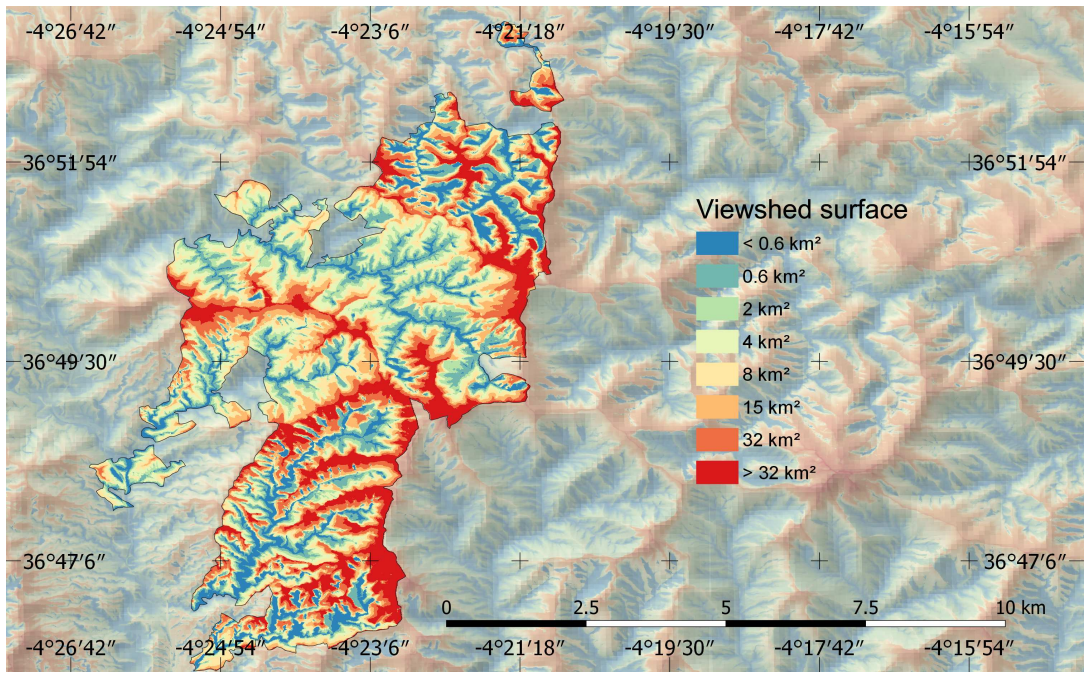}
    \caption{Total viewshed map of the Montes de Malaga Natural Park and its surroundings in the province of Malaga, Spain.}
    \label{fig:9}
\end{figure}

\subsubsection{Total viewshed map generation}
The final outcome from computing our proposed sDEM algorithm to obtain the total viewshed map of the Montes de Malaga Natural Park (Malaga, Spain) is presented in Figure~\ref{fig:9}. No substantial differences have been found after analyzing the values of absolute and relative differences when comparing the total viewshed results from the TVS and sDEM algorithms. The DEM10m-2k was used for this analysis, where the absolute difference found is up to 1.18\%, whereas the relative difference is up to 4.21\%. All these values are within the limits recommended in this field~\citep{tabik2013simultaneous}.

\section{Conclusions}\label{conclusions}
In this paper, we present a new methodology called skewed Digital Elevation Model (sDEM) to speed-up terrain surface analysis. Total viewshed computation was selected as a case study to assess the performance of this new methodology, which is designed from scratch and differs from state-of-the-art methods in the way that operations are performed. It focuses on increasing the performance of memory accesses by applying a data restructuring before starting the computation. The proposed data reorganization opens the door for intensive use of GPUs in many algorithms for which it had never been considered due to their irregularity and low efficiency.

Different versions of our algorithm have been proposed for single-core, multi-core, single-GPU and multi-GPU platforms, along with intensive performance studies compared to the literature. sDEM has been tested on Windows and Linux operating systems using two different systems and three DEMs of up to 64 millions points from the Montes de Malaga Natural Park (Malaga, Spain). Our implementations performed better than the most commonly used GIS software regarding the multiple viewshed computation. In fact, this difference would be much greater when considering all the points in the terrain but current GIS software is unable to carry out this task. Moreover, sDEM largely outperforms the state-of-the-art algorithm in terms of speed-up and throughput for the three evaluated DEMs. In particular, our approach accelerates this computation up to 827.3x for the best studied case with respect to the baseline single-threaded implementation on a given DEM formed by 16 million POVs. 

Our algorithm can be used for analyzing the surface of any terrain. For example, the computation of the viewshed map of any terrain is faster with sDEM than with the approaches reported in the literature. Also, the analysis of other topographic features such as slope and elevation could be improved by applying our methodology.

\section*{Data and codes availability statement}
The data and codes of the sDEM algorithm that support the findings of this study are available in the `figshare.com' repository with the identifier `https://doi.org/10.6084/m9.figshare.11370549'.

\section*{Funding}
This work was supported in part by the Spanish Ministry of Science and Technology through the TIN2016-80920-R Project, and in part by the University of Malaga
through the U-Smart-Drive project and the I Plan Propio de Investigacion. Siham Tabik was supported by the Ramon y Cajal Programme under Grant RYC-2015-18136.

\bibliographystyle{tfv}
\bibliography{interacttfvsample}

\end{document}